\documentclass[aps,pre,amsmath,amssymb,reprint,superscriptaddress,preprintnumbers]{revtex4-1}

\usepackage{graphicx}
\usepackage{dcolumn}
\usepackage{bm}

\usepackage[utf8]{inputenc}
\usepackage[T1]{fontenc}
\usepackage{eurosym}
\usepackage{mathptmx}
\usepackage{etoolbox}
\usepackage{amssymb}
\usepackage{array}
\usepackage{booktabs}

\usepackage{xcolor}
\usepackage[normalem]{ulem}
\usepackage{lineno}

\begin{document}

\title{The predictive power of the Blockhain transaction networks: Towards a new generation of network science market indicators} 

\author{M. Grande}
\email[Email address: ]{mar.grande@alumnos.upm.es}
\affiliation{Grupo de Sistemas Complejos; Universidad Polit\'ecnica de Madrid; 
 28035 Madrid (Spain)}
\affiliation{AgrowingData; Navarro Rodrigo 2 AT; 04001 Almer\'ia (Spain)}

\author{F. Borondo}
\email[E--mail address: ]{f.borondo@uam.es}
\affiliation{Departamento de Química; Universidad Autónoma de Madrid;
CANTOBLANCO - 28049 Madrid, Spain}

\author{J. Borondo}
\email[\textit{Corresponding author} email address: ]{jborondo@gmail.com}
\affiliation{AgrowingData; Navarro Rodrigo 2 AT; 04001 Almer\'ia (Spain)}
\affiliation{Departamento de Gesti\'on Empresarial; Universidad Pontificia de Comillas ICADE;
  Alberto Aguilera 23; 28015 Madrid (Spain)}

\date{\today}

\begin{abstract}
Currently cryptocurrencies and Decentralized Finance (DeFi), which enable financial services on public blockchains, represents a new growing trend in finance. In contrast to financial markets, ruled by traditional corporations, DeFi is completely transparent as it keeps records of all transactions that occur in the network and makes them publicly available. 
The availability of the data represents an opportunity to analyze and understand the market from the complexity that emerges from the interactions of the actors (users, bots and companies) operating in the embedded market. In this paper we focus on the Ethereum network and our main goal is to show that the properties of the underlying transaction network provide further and useful information to forecast the evolution of the market. We aim to separate the non-redundant effects of the blockchain transaction network properties from classic technical indicators and social media trends in the future price of Ethereum. To this end, we build two machine learning models to predict the future trend of the market. The first one serves as a base model and considers a set of the most relevant features according to the current scientific literature—including technical indicators and social media trends. The second model considers the features of the base model, together with the network properties computed from the transaction networks. We found that the full model outperforms the base model and can anticipate 46\% more rises in the price than the base model and 19\% more falls. Thus, we conclude that indicators based on network properties provide valuable information to forecast the future direction of the market that can not be explained neither by traditional indicators, or social media trends. Hence, our results represent a first step towards a new family of DeFi market indicators based on the complexity of the underlying transaction network.
\end{abstract}


\maketitle 

\fontsize{10}{11}
\section{Introduction}
Cryptocurrencies and Decentralized Finance (DeFi), an emerging financial technology based on secure distributed ledgers, represent a new trend in finance that is growing exponentially, taking components of traditional financial markets and transforming them into transparent and decentralized protocols through smart contracts and tokens. 

The appeal that this new paradigm offers to the population is clear: access to all imaginable financial services ---savings accounts, insurance, loans, commerce--- in an open and borderless way. In this new scenario, all financial services are carried out through decentralized applications that run on the blockchain and therefore eliminate the need for a central authority, which is its differential advantage over traditional markets. Moreover, during the last decade the cryptocurrencies and DeFi market has become the fastest-growing alternative investment option, gaining widespread public and investors attention  due to their extraordinary returns in phases of extreme price growth. Another  differential advantage of this new market for investors is that, because of its decentralized nature, it does not react to economic factors.  Thus, the failure of governments and central banks in the 2008 financial crisis alongside the European Sovereign Debt Crisis, represented an incentive for investors and other economic actors towards cryptocurrencies that began to see them as an attractive investment alternative. However,  the cryptocurrency market is still in the nascent stages which involves a high  volatility,  presenting remarkable fluctuations and unexpected massive crashes. 

The dynamics of the cryptocurrency market have attracted the interest of financial institutions and researchers seeking to understand its high volatility and price fluctuations. The behavior  and evolution of all financial markets depends largely on the behaviors of traders  and investors operating in such markets, and the cryptocurrencies and DeFi markets are no exception. Seeking to take advantage of this behavior researchers developed over the past century Technical Analysis (TA) \cite{park2004profitability, brown1989technical, treynor1985defense, park2004profitability}, a methodology to analyze and forecast the direction of prices through the study of past market data (only  considering price and volume). This theory is founded on the principle that a market's price time series reflects all relevant information impacting that market. Thus, researchers in this field have developed over the past decades TA  indicators based on the history of an asset or commodity's trading patterns. The most well-known types of systems are moving averages, channels (support and resistance) and momentum oscillators. Several studies have already applied TA to study and forecast cryptocurrencies price series, focusing mainly in Bitcoin and Ethereum, and finding a correlation between some technical indicators and their future price \cite{gradojevic2023forecasting, ilham2022effect, detzel2021learning}. More recently, other scholars have highlighted the power of social media to capture the sentiment and behavior of users. Thus, a recent line of research focuses on generating indicators based on social media trends to forecast financial markets. Several studies have focused on exploring the relation between social media and the price of stocks  \cite{nguyen2015sentiment, sun2016trade}, while more recently, other researchers have focused on the relation of these indicators towards the cryptocurrency market \cite{lamon2017cryptocurrency, phillips2017predicting,ortu2022technical}. 

However, a differential factor of this new system with respect to traditional markets is that it keeps records of all transactions that occur in the network and makes them publicly available. The open nature of blockchain transactions involves an opportunity and need to research new tools and methods to analyze and track the movements of this novel market. In contrast to other financial markets, we can process all the financial interactions taking place on public blockchains, and model the system as a network \cite{barabasi2013network, barabasi2002new,newman2011structure}, analyzing it from the perspective of complex systems \cite{boccaletti2006complex, strogatz2001exploring, foster2005simplistic}. This approach enables us to capture the flow of information in the underlying network and track the global behavior that emerges from the interactions between users. Hence, the network representation and analysis of these publicly available transactions enables a new form of financial econometrics—with the emphasis on the complexity that emerges from the interactions of users rather than solely the covariance of historical time series of prices and volume, how TA does. In this work, we take advantage of network science to characterize the market, and track and understand its evolution over time. 

\begin{figure*}
\includegraphics[width=1.0\textwidth]{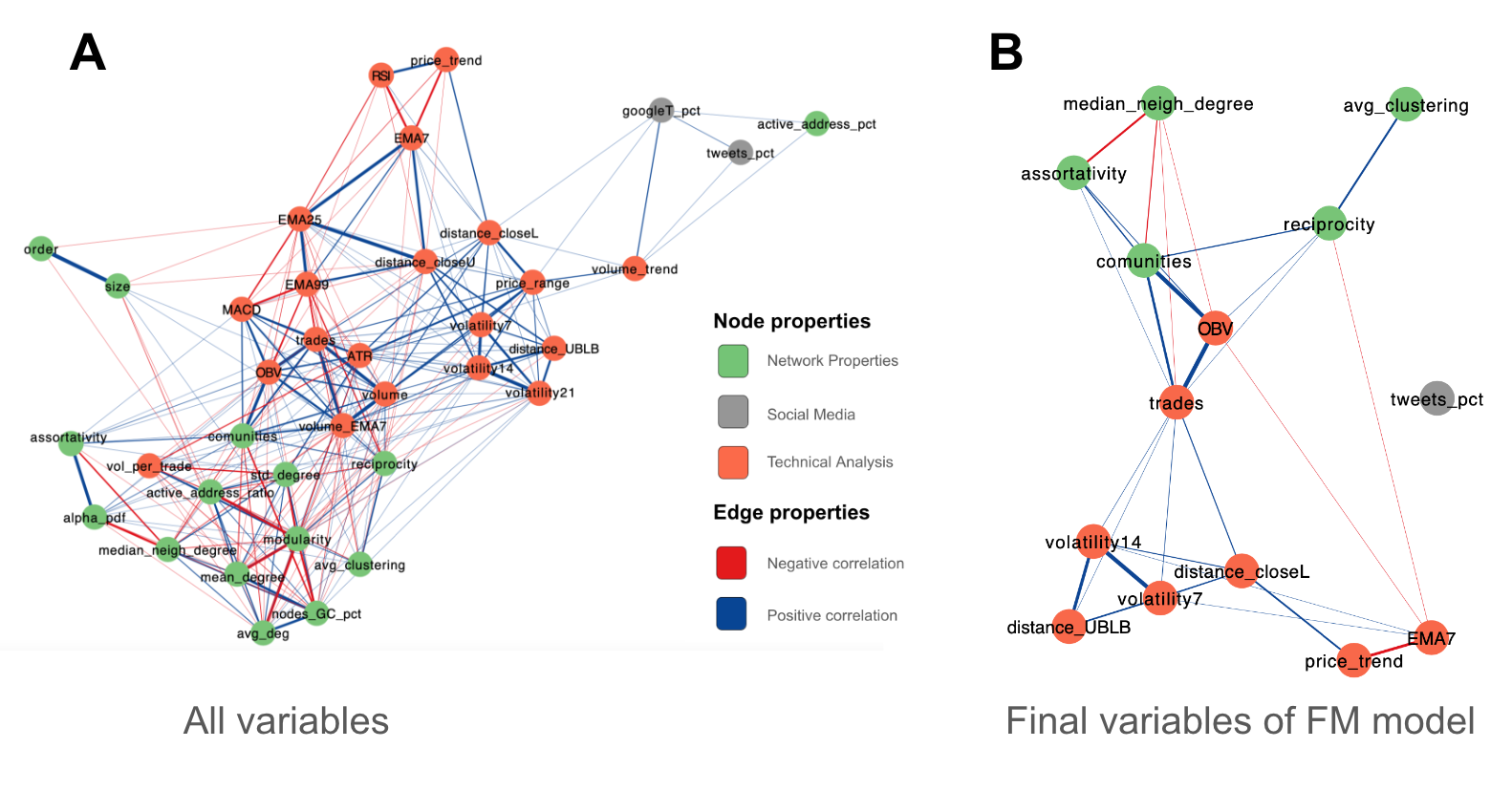}
\caption{This figure summarizes the relations between the variables of the three family of indicators: network properties, technical analysis and social media. Panel A show the network resulting from the correlation between all the variables considered in the study. Panel B shows a filtered version of the network, only including the variables selected by the final model. Note that only significant correlation are represented as links ($\text{p-value}<0.05$)}
\label{fig:network}
\end{figure*}

Recent scientific contributions have analyzed the transaction network of several cryptocurrencies  \cite{vallarano2020bitcoin}. In \cite{LiangBTC} the authors characterize the properties and the temporal evolution of the transaction network of Bitcoin, Ethereum and Namecoin, finding significant variations in their structural properties over time. Other works have gone one step further, exploring the  relationship of network properties with the price, and finding evidence to support that some subgraphs of the transaction network can influence Bitcoin price \cite{AKCORA2018138}. In the same line other authors have identified a relation between certain network properties an bitcoin bubbles \cite{bovet2018network}. All of these works have focused on analyzing the transaction network separately and not controlling for other factors that could provide similar information.

But neither social media trends nor TA indicators  are independent from the  network of transactions. All three families of indicators seek to capture the behavior of users and are therefore related. In fact, all the indicators form a network, where they tend to cluster according to their family, but with clear relations across the three sets (see Figure \ref{fig:network}). Hence, to properly interpret relationships between the evolution of the price of cryptocurrencies, and the behavior of users operating in the market, one needs to separate the effects of the transaction network properties from the social media trends and the information already captured by TA. By separating these effects we can claim whether the network of transactions contains relevant  and novel information about the market or not.

This paper aims to contribute to understanding the crypto and DeFi markets by taking a first step towards a new line of research in financial econometrics based on the complexity that arises from the interactions that take place in public blockchains (between billions of individuals, entities and bots). In this work, we focus on Ethereum (ETH) because it is by far the largest blockchain supporting smart contracts  and the majority of DeFi protocols and applications are built on it. We puzzle out the effect that the transaction network properties, TA indicators and social media trends have on the future  price of ETH. To this end, we build two machine learning models to predict the future trend of ETH. The first one serves as a base model and considers a set of the most relevant features according to the current scientific literature—including TA indicators and social media trends. The second model considers the features of the base model, together with the network properties computed from the transaction networks. We find that even after controlling for TA and social media, the network properties still provide valuable information to forecast the direction of the price, as the Full Model outperforms the Base Model. In addition, the variables related to network properties have a significant effect in the model. Finally, we argue that our results highlight the need to develop a new family of network based indicators for the emerging cryptocurrencies market.

\section{Results: The predictive power of the ETH network}
\label{Results}

We begin by developing a machine learning model, based on the XGBoost algorithm, with the categorized logarithmic return of the close price as the dependent variable. The logarithmic return is defined as: 

$$r_{t+1} = \ln(\frac{p_{t+1}}{p_t})$$

where $p$ is the close price at a timestamp $t$. 

Next, to categorize the target, we consider that the logarithmic return has increased (decreased) within consecutive days if the logarithmic return change is greater than 1\% (lower than -1\%). Otherwise, we say that the logarithmic return has remained constant. Hence, the target function takes a value of 1 to indicate an uptrend, -1 to indicate a downtrend, and 0 when it remains constant.
As explanatory factors we use three different sets of variables explained in section~\ref{featuares}. The first set of variables, related to  TA, is formed by some trend, volume, volatility and momentum indicators, which are widely used in the scientific literature to anticipate the evolution of markets. The second set of variables, related to social media trends (SM), is formed by the variables \textit{Number of Tweets} and \textit{GoogleTrends}. This group accounts for the information or impact that social media activity can have on the future prices. Finally, the third set of variables, related to the network properties (NP), is formed by the main size, degree and connectivity properties of the network. Full details of all the variables build and their description can be fount in section~\ref{featuares}. With these three sets of variables, we set up two different specifications of the ML model. The first specification, or Base Model (BM), serves as a baseline and includes the TA variables plus the SM variables. The second specification of the model, or Full Model (FM), adds the NP variables to the BM model.

\begin{figure*}
\includegraphics[width=1\textwidth]{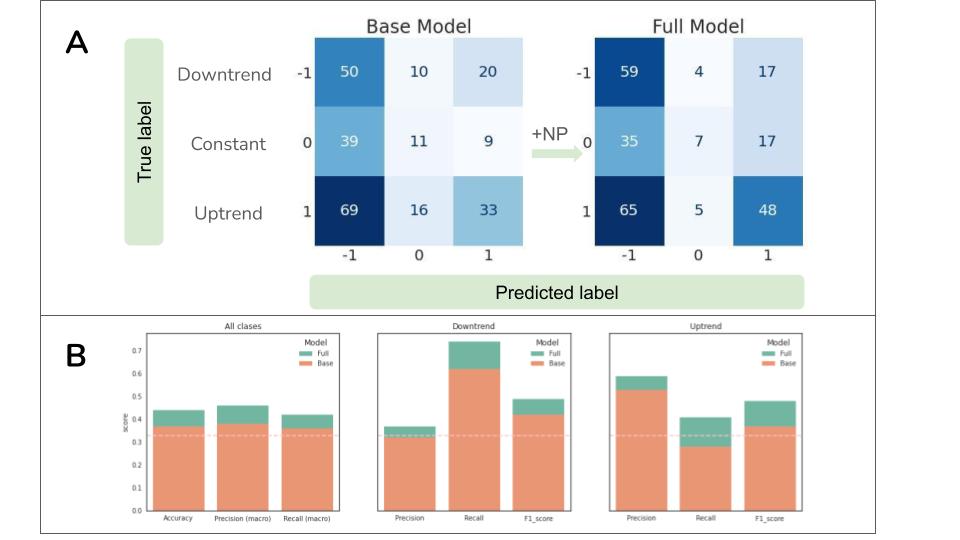}
\caption{This figure compares the performance of the BM  and FM models. Panel A shows a comparison between the confusion matrix of both models, reflecting that the FM model achieves a higher accuracy. Panel B compares the precision, recall and F1-score of both models for all classed, uptrends and downtrends. As it can be seen the FM model outperforms in all cases.}
\label{fig:results}
\end{figure*}

The results obtained by both models are summarized in Figure \ref{fig:results}. In panel A of this figure we compare the confusion matrix of both models. As it can be seen the FM model achieves a higher accuracy. To provide extra detail panel B of the figure compares the precision, recall and F1-score of both models for all classes, uptrends and downtrends.

The BM model has an accuracy of 37\%, a precision of 38\% and a recall of 36\%. Analyzing the model in further detail we show that, anticipating downtrends, the Base Model shows poor precision (32\%) but a reasonably high recall (62\%). This means that the model is capable of anticipating 62\% of the future price drops, but in counterpart only 32\% of the downtrend signals will actually happen. Regarding uptrends, the BM model exhibits the opposite behavior.  The recall of the model is poor (28\%), while the precision increases up to 52\%. Thus, the model only anticipates 28\% of the actual uptrends, but over half of the times this signal is actually true.
Next, we analyze the results of the FM model, which actually significantly outperforms the BM model. This model achieves an accuracy of 44\% (19\% higher than BM), precision of 46\% (21\% higher than BM), and recall of 42\% (17\% higher than BM). Regarding the performance of the model on anticipating uptrends, the FM model increased the up recall on 46\% points in comparison to the BM model, reaching a final recall of 41\%. At the same time, the FM model obtained a higher precision (59\%) in uptrends than the BM model. This means that the FM model anticipates 41\% of the total uptrends, being almost 60\% of these forecasts actually true. Regarding downtrends, the improvement of the FM is slightly smaller. However, still the FM has 16\% higher precision ($P_{FM}=37\%$) and 19\% higher recall ($R_{FM}=74\%$). This means that despite the model not being very precise anticipating downtrends, it is capable of alerting almost 3 out of 4 of price drops.
Finally, we analyze the final features selected and their corresponding importance in the best configuration of the FM model.  We observe that network properties play an important role in this final best model,  where from the final 15 variables 6 of them are generated from the network properties. This is illustrated in Figure \ref{fig:network}, that shows in panel A a network representation of all the initially considered variables, and in panel B, the final variables selected in the FM. In addition,  these variables in general occupy top position in the ranking of feature importances according to different metrics, including cover, gain, or mean decrease impurity among others (see the Methods section for a full explanation on these metrics, and Appendix for a table with the full results). We also have to highlight that all three families of variables are present in the FM model. Hence, the good performance of the model, that clearly outperforms the BM model and random guessing, results from the combination of the information hidden in the price time series, social media and the network of transactions.

\section{Discussion}
  \label{Discussion}
Today's society is evolving towards a digital society where the adoption of cryptocurrencies among citizens and companies grows every day, being used to trade, invest and speculate, as well as to access services. In fact, today, there are more than 1,500 cryptocurrencies and according to a recent study~\cite{rauchs20192nd} it is estimated that more than 35 million of private and institutional investors participate in the different transaction networks. This means that in order to anticipate new economic crises, as well as promote a fair economy, it is necessary to know and understand the mechanisms that move the cryptocurrency market. 

One of the key features of this emerging market  that differentiates it from traditional financial institutions is that it operates on decentralized networks with verifiable transactions that are publicly available. This availability of the transactions opens an opportunity to analyze the market from a novel perspective. Thus, we can take advantage of network science \cite{barabasi2013network, barabasi2002new,newman2011structure} to capture the complex global behavior that emerges from the individual transactions taking place in the blockchain, and that drives the market. In this work, we have analyzed the evolution of properties of the transaction network, and shown that the information generated from them is useful to anticipate the evolution of the prices even after controlling by TA and social media trends. In fact, this represents one of the main contributions of our paper. In contrast to previous studies that have focused on analyzing the effect of each family of indicators separately, we analyze them together untangling their impact in the evolution of prices.

The fact that the properties of the transaction network provide additional information that is not captured by TA indicators, demonstrates that the transaction history includes extra information that can not be obtained solely from the analyzing the time series of prices and volumes. Hence, this result has important implications for the Efficient Market Hypothesis (EMH) debate \cite{fama1970efficient,fama1991efficient}. The EMH, was introduced  by Fama in 1970, and states that new information is immediately reflected in asset prices, that therefore show martingale behavior. Thus this scenario implies that no trading rules based on the market time series can obtain returns that surpass the buy-and-hold strategy. The current literature is full of papers both supporting and rejecting the EMH \cite{greene1977long, lillo2004long,jacobsen1996long, lo2011non, di2005long} in several markets. However, since most of these works focus on markets where transactions are not public it has not yet been analyzed or discussed the implications that the network representation of the transactions have on the predictability of the market.

But do our results mean that the information generated from the complexity that emerges from the transaction network is enough to understand the market? Not really. Under the light of our results, the information generated from the underlying transaction network is more valuable when combined with the price time series and social media trends. In fact, we have shown that these three sources of information are related but at the same time contain complementary information. Hence, we argue that the new cryptocurrency market, where transparency and the public nature of transactions are  key factors, opens a new era in financial econometrics. Thus, opening a new line of research focused on developing new methodologies and indicators that capture the complexity that emerges from the transaction between users.

Finally, the fact that network properties provide new information to understand the cryptocurrency markets invites us to explore further and more sophisticated network representations and properties. In this regard, in this paper we have built a single layer representation of the Ethereum system. However, there are multiple applications with their own token running on top of the Ethereum chain. Hence, it would be of particular interest to model the system as a multiplex network \cite{BOCCALETTI20141,battiston2017new, gomez2013diffusion}, where the nodes (wallets) can interact through multiple layers, each one representing a different application and token. In addition, the features built from the network properties can be expanded in the future. For this work, we have only included basic network properties to show the potential of this line of research. However, once that the usefulness of this information is clear, more research should be conducted in order to build more informative indicators.

\section{Materials and Methods}
  \label{Methods}

\begin{figure*}
\includegraphics[width=0.5\textwidth]{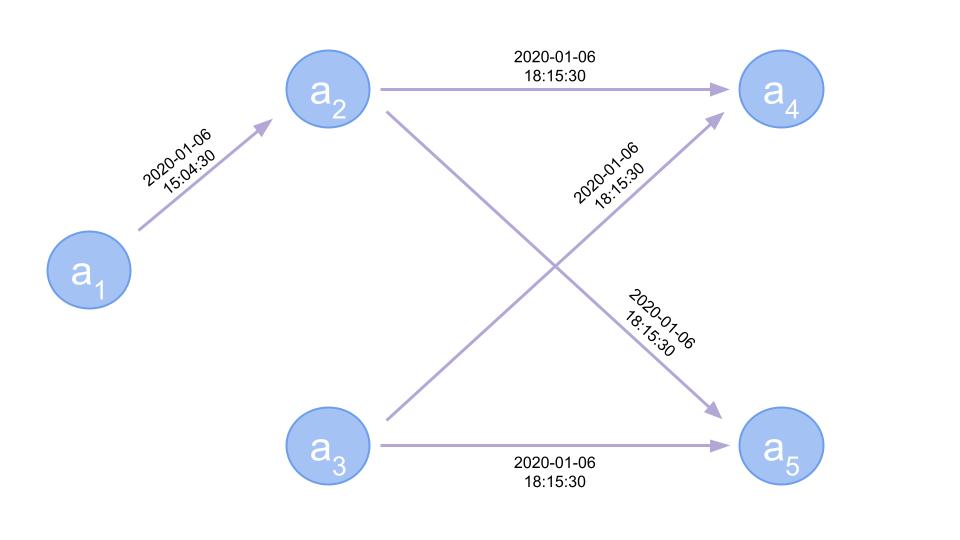}
\caption{Network representation of two transactions in the Ethereum system. Nodes represent addresses and the links ---denoted with arrows--- represent transactions between two addresses. The property of the edges is the timestamp in which transactions were added to the Blockchain.}
\label{fig:transaction_example}
\end{figure*}

\subsection{Dataset}
  \label{Dataset}
In the present work we analyze the network of transactions that take place on the Ethereum Blockchain. These transactions are public and real time available through \url{https://etherscan.io}. We collected all the transactions generated from January 2018 to January 2021, resulting in more than 3 billion transactions.

The second part of this work focuseds on analyzing the relation between the properties of the transactions network and the price of ETH. For this, we consider a second dataset that includes the time series of the daily price. This second dataset was collected from Binance exchange, using its API. From the data sollected from Binance we computed the most widely used TA indicators that have already reported good results in current literature for price prediction problems. Finally, we include a third family of variables in our analysis related to social media trends. To this end we obtained and computed the number of tweets mentioning Ethereum and google trend score for Ethereum.

\subsection{Ethereum transaction network}
  \label{eth_network}
The Ethereum system is an open source infrastructure built on and supported by a blockchain, i.e. a decentralized public ledger which records all transactions that occur in the system. Once these transactions are uploaded to Ethereum, they cannot be deleted or modified, which makes the system reliable. 
Each transaction consists of a source address (where currencies come from), a target address (where currencies go), the amount of transferred coins and the timestamp among other attributes. 

In this paper, we built the transaction network between addresses, in which nodes represent the addresses that operate in the system and links represent the transactions between them. The direction of links is determined by the source and target addresses and the weights are defined as the number of transactions between each pair of nodes for each day. An example of the network construction is illustrated in Fig.~\ref{fig:transaction_example}.

\begin{figure*}
\includegraphics[width=1.0\textwidth]{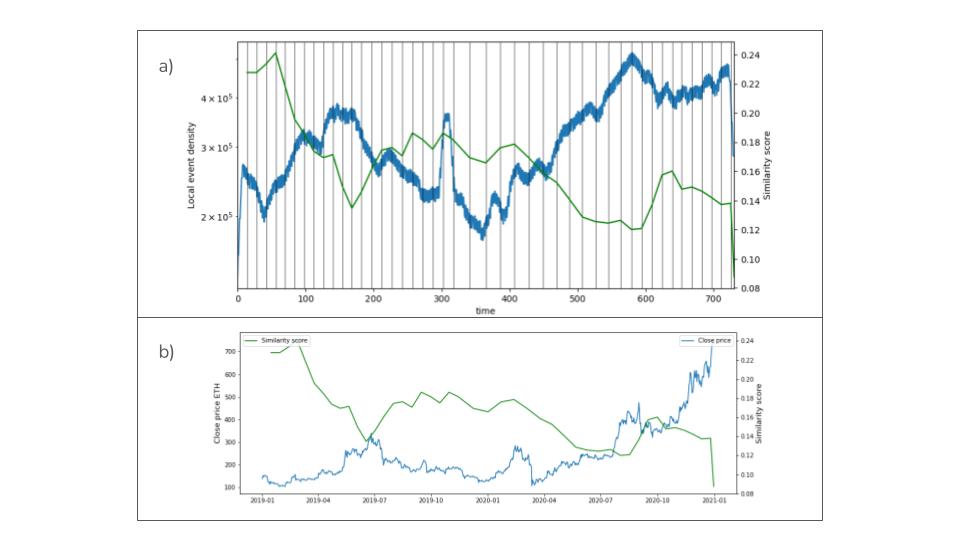}
\caption{Panel a) shows the time intervals (vertical lines) and similarity score (green lines) computed with Dynamic time-slicing method. The blue line (y-axis) represents the average number of weighted events by the number of transactions. Panel b) shows the evolution of the similarity score (green line) with the close price of Ethereum (blue line).}
\label{fig:sim_events}
\end{figure*}

\subsubsection{Time aggregation}
  \label{time_agg}
The Ethereum transaction network is an evolving complex system. Nodes are added to the network when new addresses are created and removed when they are no longer involved in any transaction, while new edges arise when there is a transaction between two previously unconnected addresses. Therefore, in order to understand the dynamics of this system, we need to define dynamic intervals that match its evolution. The choice of these intervals is key in this work because they will condition our future analysis. In contrast to other works, where the time aggregation is fixed and randomly chosen~\cite{elbahrawy2017evolutionary}, we apply the method developed by Darst et al.~\cite{darst2016detection} that detects evolutionary changes in the configuration of a complex system and generates intervals accordingly. The size of each interval is determined by maximizing the similarity between the sets of events within consecutive intervals. In this case, the events are the transactions recorded in the Ethereum Blockchain and the similarity metric used is the jaccard index. When applying this method to our dataset, we found that the optimal intervals obtained are regular ---around 14 days. Figure ~\ref{fig:sim_events}a) shows a visualization of these intervals, along with the similarity score obtained. 

In Figure~\ref{fig:sim_events} b) we compare the evolution of the obtained similarity score with the time series of the price observing a strong and  significant relation, proxied as the Pearson correlation coefficient (r=-0.67, p-value=3.76e-07 ) with the price. In summary, we found that the higher the price, the lower the similarity score. As a lower similarity implies higher density of events, this result means that there is a positive linear relation between the number of events ---i.e., number of transactions--- and the price. 

\subsection{Model features}
  \label{featuares}
In this section we describe the construction of the three sets of variables we use to predict the future trend of ETH price. To simplify the presentation, we divide our twenty five variables into three groups: network properties indicators (NP), TA indicators and social media trends (SM).

The first set describes some important features that capture the overall structure of the network giving insights into network properties such as its connectivity. Some features are individual properties of each node, such as  the \textit{degree} or \textit{PageRank}. In these cases, we use the mean and standard deviation of these values as our topological feature for the entire network. 

The second set of features considers the most relevant TA indicators according to the current scientific literature for price prediction problems ---including trend, volatility, volume and momentum indicators.

The third set is composed by two variables that gather information about social media trends such as the number of tweets mentioning Ethereum and the score in Google Trends. 
\subsubsection{Network properties}
  \label{net_feat}

We compute the main properties of the network such as the number of nodes and links of each network and the degree of each node and its neighbors. Taking into account the degree distribution of the nodes, we also build as a feature the slope of this distribution in a log-log scale (the alpha coefficient if it can be approximate by a power-law distribution). We also include the ratio of active addresses (nodes in the network) versus total addresses that there are in the system.

Other network properties that we included regarding the connectivity of the network are:
\begin{itemize}
\item Assortativity. In this work, we focus on the assortativity by degree, i.e., the tendency of nodes to connect others with similar degrees. We compute the assortativity of the network according to the degree correlation coefficient introduced by Newman\cite{newman2002assortative}. The formula is as follows:
$$r=\sum_{jk}\frac{jk(e_{jk}-q_{j}q_{k})}{r^2}$$
with
$$r^2=\sum_{k}k^{2}q_{k}-\left[\sum_{k}^{jk}kq_{c}\right]^2$$
where $e_{jk}$ is the probability of finding a link between the nodes $j$ and $k$ and $q_{k}$ the probability that there is a degree-$k$ node at the end of the randomly selected link. 

\item Clustering coefficient. It measures the fraction of possible triangles that are actually closed. In the context of transaction networks, a high clustering indicates that people to whom I make transfers, make transfers between them. On the other hand, a low clustering means that my neighbors do not transfer currencies between them. In this work we calculate the local clustering of nodes according to the method proposed by Watts \& Strogatz\cite{watts1998collective}. Thus, for a node of degree $k_{i}$ the local clustering can be expressed as:
$$C_{i} = \frac{2L_{i}}{k_{i}(k_{i}-1)}$$
where $L_{i}$ represents the number of links among the neighbors of node $i$. Hence, the average clustering coefficient of the network can be calculated as the average of $C_{i}$ for all nodes:
$$\langle C_{i}\rangle=\frac{1}{N}\sum_{i=1}^{N}C_{i}$$

\item Communities and modularity. A community can be defined as a group of nodes that have a higher likelihood of connecting to each other than to nodes from others communities. The quality of a particular partition of a network into communities can be measured by the modularity. It can be formulated as follows: 
$$Q=\frac{|E_{in}|-\langle |E_{in}| \rangle}{|E|} $$
where $|E|$ is the number of edges, $|E_{in}|$ is the number of edges within the community and $\langle |E_{in}| \rangle$ is the expected number of edges within the community if the topology were purely random. It ranges between $[-1, 1]$, where 1 indicates there are no links across different communities and -1 indicates that there are no links between the nodes of a community. In this work, we use the \textit{Louvain} method proposed by Vincent Blondel et al.\cite{blondel2008fast} in 2008. This method iteratively optimizes local communities until global modularity can no longer be improved given perturbations to the current community state. 

\item Reciprocity. It is a metric for directed networks that measures how likely is that a node you point to also points back at you. More concretely, if there is an edge from node $i$ to node $j$ and there is also an edge from $j$ to $i$ then we say the edges are reciprocated. The reciprocity $r$ is defined as the fraction of edges that are reciprocated.

\item PageRank (PR). It is a centrality measure that determines the relative importance of a node within the network. A node has a high PageRank when it is highly connected or when it is attached to leading ones. The PageRank value for any node $u$ can be expressed as:
$$PR(u)=\sum_{v \in B_{u}}\frac{PR(v)}{L(v)}$$
thus, the PageRank value for a node $u$ depends on the PageRank of its neighbors ($B_{u}$) and the number of links of its neighbors $L(v)$.

\item Largest connected component (LCC). It is a maximal subgraph in which any two nodes are connected by a path. In this work, we did not take into account the direction of the edges in order to compute the LCC. LCC is an important factor in understanding the network structure and can also be related to price. For instance, we expect that the higher the number of nodes in the LCC, the higher activity in the network, which would lead to fluctuations in the price. 

\end{itemize}

\subsubsection{Technical analysis indicators}
  \label{ta_feat}
\begin{itemize}
    \item Price trend. It is defined as the difference between the close and open price.
    \item Exponential moving average of close price (EMA). We consider three times windows for this feature: 7, 25 and 99 days.
    \item Moving average convergence divergence (MACD). It is a trend-following momentum indicator that shows the relationship between two moving averages of the price. It is calculated as the difference between the EMA in a window of 26 periods and the EMA in a window of 12 periods. A 9 periods EMA of the MACD is considered as the signal line, which serves as the threshold for the buy or sell signals.
    \item Relative strength index (RSI). It is a popular momentum indicator which determines whether the stock is over-purchased or over-sold.
    \item Volume. Total volume traded during the day. We also consider the trend of the volume define as the change in percentage with respect to the volume of the previous day.
    \item On balance volume (OBV). It is a technical indicator used to find buying and selling trends of a stock, by considering the cumulative volume: it cumulatively adds the volumes on days when the prices go up, and subtracts the volume on the days when prices go down, compared to the prices of the previous day.
    \item Volatility. It is defined as the standard deviation of the logarithmic return of the close price. We consider a time window of 7, 14 and 21 days in order to build this feature.
    \item Average true range (ATR). It is a measure of the market volatility and it is computed as follows: 
    $$ATR=\left( \frac{1}{n}\right)\sum_{i=1}^nTR_{i}$$
    with
    \begin{equation}
    TR_{i} = \max
    \begin{cases}
      \text{high}_i - \text{low}_i \\
      |\text{high}_i - \text{close}_{i-1}| \\
      |\text{low}_i - \text{close}_{i-1}|
    \end{cases}       
\end{equation}
    where $TR_{i}$ is a particular True Range and $n$ is a time period.
    \item Price range. It is defined as the difference between the high price and the low price for each day.
    \item Bollinger Bands. It is composed by a set of trend lines. The central band is calculated from a simple moving average (SMA) of the close price. The upper and lower bands are then calculated by adding and subtracting two times the standard deviation of the central band respectively. The time window used to compute the SMA is 21 days. As features for the model we calculate the distance in percentage between the upper and lower band and between the close price and each band. 

\subsubsection{Social media trends}
  \label{sm_feat}
    \item Number of tweets. It is the number of tweets using \#Ethereum hashtag in Twitter for each day. For this feature, we compute the change in percentage with respect to the previous day. 
    \item Google Trend score. It is related to the number of searches in Google mentioning Ethereum. For this feature, we compute the change in percentage with respect to the previous day. 
\end{itemize}

\subsection{Machine Learning models}
\label{MLmodel}
In this work, we use three families of features to forecast the future trend of the ETH, price and are not only interested in the predictions, but also in understanding the contribution of each family of features. In this scenario we have decided to use machine learning (ML) as it can be successfully applied to financial problems \cite{dixon2020machine,de2018advances}.  We believe ML is the appropiate decision mainly because	we are dealing with  high-dimensional data where the features are interrelated. Thus, the flexibility of ML in comparison to classical econometric models makes them a more efficient option to identify complex patterns in a high-dimensional space.

Tree based models are popular machine learning approaches which can be used to solve a wide range of regression and classification problems. These models combined with an ensemble technique have also reported better performance than other machine learning algorithms in real world problems, like stock market forecasting. For instance, in ~\cite{dey2016forecasting} the authors proposed a system using XGBoost as a classifier in order to forecast the stock market in 60-day and 90-day periods and concluded that XGBoost turned out to be better than other non-ensemble algorithms, such as Support Vector Machines (SVM) and Artificial Neural Networks (ANN).

In this work, we train an XGBoost (Extreme Gradient Boosting)~\cite{chen2016xgboost} algorithm to predict the trend of the logarithmic return of the price. This algorithm is based on Tree Boosting, a machine learning technique that attempts to create a strong learner from a given number of weak learners, that only performs slightly better than random guessing~\cite{freund1999short}. The main principle of boosting is to iteratively fit a sequence of weak learners to weighted versions of the training data. After each iteration, misclassified input data gain a higher weight and examples that are classified correctly lose weight. Thus, future weak learners focus more on the examples that previous weak learners misclassified. At the end of the process, all of the successive models are weighted according to their performance and the outputs are combined using voting for classification problems or averaging for regression problems, creating the final model.

Once the XGBoost algorithm is trained it is possible to measure the implication of each feature in the prediction. For instance, we can rank the features by the number of times they appear in a tree or by the number of predictions in which each feature participates. In this work, we use this criteria in order to measure the contribution of each variable in the model and select a final set of features that provide the best prediction results.

\subsubsection{Training and testing}
\label{TrainingTesting}
For training the model, we split the dataset into train and test sets. The training period starts in January 2018 and ends in June 2020, then starts the test period. Thus, there are 897 data points in the training set and 258 data points in the testing set. The training set is used to select the optimal hyperparameters of each model, so the test set is then used to evaluate the performance of each model on unseen data. 

\subsubsection{Hyperparameters}
\label{Hyperparameters}
In order to mitigate variability in the results, we have  optimized the hyperparameters for the XGBoost models  by cross-validated grid search, maximizing the area under the ROC curve of the training set. The grid search is composed by the following parameters:
\begin{itemize}
    \item \textit{n\_estimators}. It is the number of trees in the algorithm.
    \item \textit{max\_depth}. It is the maximum depth that a tree can grow.
    \item \textit{colsample\_bytree}. It is the subsample ratio of columns when constructing each tree. Subsampling occurs once for every tree constructed.
    \item \textit{min\_child\_weight}. It is the minimum sum of instance weight needed in a child. If the tree partition step results in a leaf node with the sum of instance weight less than \textit{min\_child\_weight}, then the building process will give up further partitioning.
    \item \textit{learning\_rate}. It is the step size shrinkage used in update to prevents overfitting. After each boosting step, we can directly get the weights of new features, and learning rate shrinks the feature weights to make the boosting process more conservative.
    \item \textit{gamma}. It is the minimum loss reduction required to make a further partition on a leaf node of the tree.
    \item \textit{subsample}. Subsample ratio of the training instances. Setting it to 0.5 means that XGBoost would randomly sample half of the training data prior to growing trees. and this will prevent overfitting. Subsampling will occur once in every boosting iteration.
\end{itemize}

The optimal values obtained in this way for each model can be seen in table~\ref{tab:hyperparams}.

\subsubsection{Feature selection}
\label{Featureselection}
Feature selection is the process of selecting a subset of variables from the input data that reduce effects from noise or irrelevant variables and still provide good prediction results. This technique is widely used in machine learning problems because it helps to reduce the overfitting and complexity of the model while improving the accuracy.

In this work we take into account the importance of each feature provided by the XGBoost algorithm in order to select the best set of features. 

\begin{table}
    \centering
    \begin{tabular}{ccccccc}
    \hline\hline \\
        Parameter                    & Base Model & Full Model \\  
    \hline \\
       n\_estimators                    & 100     & 100  \\
       max\_depth                    & 3     & 5   \\
       colsample\_bytree         & 0.5     & 0.5   \\
       min\_child\_weight & 13     & 13   \\
       learning\_rate    & 0.05     & 0.01   \\
       gamma             & 0.08       & 0.05   \\ 
       subsample        & 0.5        & 0.5   \\  
       \hline
    \end{tabular}
    \caption{Optimal hyperparameters used to train the machine learning models.}
    \label{tab:hyperparams}
\end{table}

\section{Competing interests}
The authors declare no competing financial or non-financial interests.

\section{Author contributions}
All authors developed the idea and the theory. MG performed the calculations and analyzed the data. All authors contributed to the discussions and interpretations of the results and wrote the manuscript.

\begin{acknowledgments}
The project that gave rise to these results received the
support DGof Research and Technological Innovation of
the Comunidad de Madrid (Spain) under Contract No.
IND2022/TIC-23716
\end{acknowledgments}

\bibliography{eth}

\clearpage
\newpage

\section{Appendix}

\subsection{Features importances }
\begin{figure*}
\includegraphics[width=0.5\textwidth]{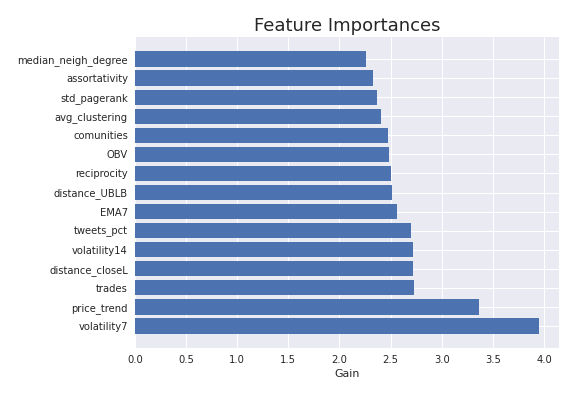}
\caption{Feature importances measure by the average gain (improvement in the score) of splits which use the feature.}
\label{fig:gain}
\end{figure*}

\begin{figure*}
\includegraphics[width=0.5\textwidth]{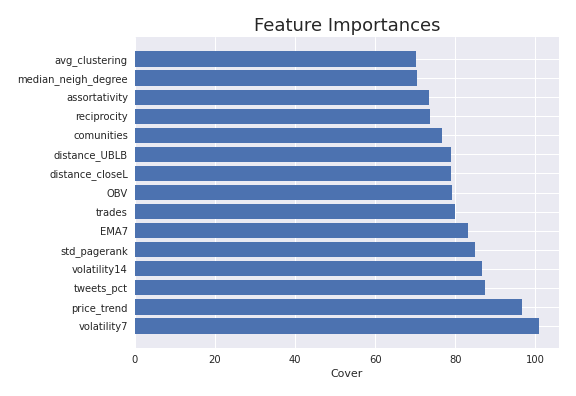}
\caption{Feature importances measure by the average coverage of splits which use the feature.}
\label{fig:cover}
\end{figure*}

\begin{figure*}
\includegraphics[width=0.5\textwidth]{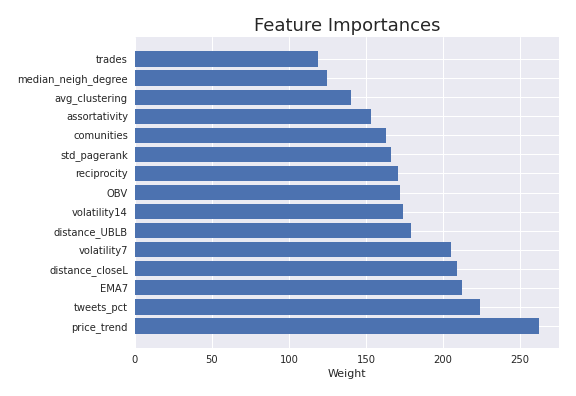}
\caption{Feature importances measure by the number of times each feature appears in a tree in the FM.}
\label{fig:weight}
\end{figure*}

Once an XGBoost is trained it is possible to measure the implication of each feature in the prediction. For instance, we can rank the features by the number of times they appear in a tree or by the number of predictions in which each feature participates. We use the following criteria in order to measure the contribution of each feature in the model:

\begin{itemize}
    \item Gain, it is the average gain (improvement in the score) of splits which use the feature. A higher value of this metric when compared to another feature implies it is more important for generating a prediction.
    \item Cover, it is the average coverage of splits which use the feature where coverage is defined as the number of samples affected by the split. For example, if we have 50 observations, 3 features and 2 trees, and suppose feature1 is used to decide the leaf node for 8 and 4 observations in tree1 and tree2 respectively; then the coverage for this feature is 8 + 4 = 12 observations. Thus, the cover of this feature is 12 divided by the total number of observations.
    \item Weight, it is the number of times a feature appears in a tree. In the above example, if feature$_{1}$ occurred in 3 splits and 2 splits in tree$_{1}$ and tree$_{2}$; then the weight for feature1 will be 3 + 2 = 5.
\end{itemize}
In Figures~\ref{fig:gain},\ref{fig:cover} and \ref{fig:weight} we show the contribution of each feature in the FM regarding gain, cover and weight respectively.

\begin{figure*}
\includegraphics[width=0.5\textwidth]{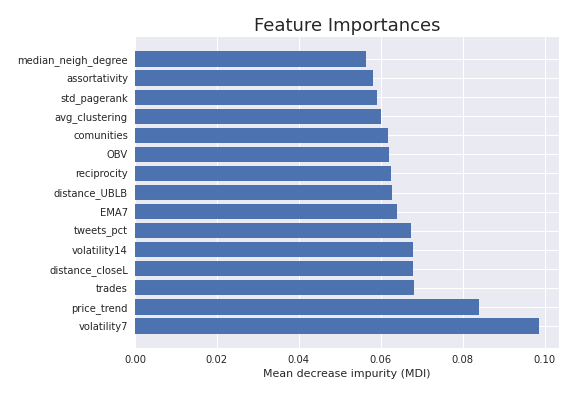}
\caption{Feature importances measure by the mean impurity decrease.}
\label{fig:mdi}
\end{figure*}

\begin{figure*}
\includegraphics[width=0.5\textwidth]{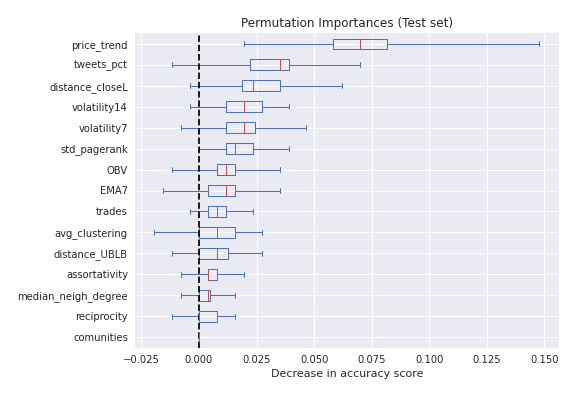}
\caption{Feature importances measure by the mean accuracy decrease.}
\label{fig:mda}
\end{figure*}

We also perform a permutation importance analysis. The full procedure essentially consists in ‘eliminating’ variables, one at a time, by making them random, and simultaneously monitoring the loss of in a model score. Here we use the accuracy score and the impurity. Figures ~\ref{fig:mda} and \ref{fig:mdi} show the results for each feature.

\end{document}